\documentstyle[osa,preprint,psfig]{revtex}
\tightenlines
\begin{document}
\title{Route to Chaos in the $^{15}NH_{3}$ far infrared ring laser}
\author{R. Dykstra, M.Y. Li and N.R. Heckenberg \\
	Department of Physics, \\
	The University of Queensland, \\
	St Lucia, QLD. 4072.}

\maketitle

\begin{abstract}

The route to chaos as the pump power is raised in  the Lorenz-like
$^{15}NH_{3}$ laser is studied using transients and compared with the route
which is expected in the Lorenz Haken equations. The differences between the
routes to chaos, as normally displayed by the experimental system and the
Lorenz Haken equations, is described in terms of the bifurcations experienced
in each system. It is also shown that the experimental system has a small
parameter regime where the route to chaos is the same as in the Lorenz Haken
equations.  \end{abstract}

\section{INTRODUCTION}

In 1963 Lorenz\cite{Lorenz:1963} reduced the equations for convective fluid
flow into three first order coupled nonlinear differential equations and
demonstrated with these the idea of sensitive dependence upon initial
conditions and chaos. It was shown by Haken in 1975 that these equations are
isomorphic with the Maxwell Bloch laser equations for a homogeneously broadened
single mode travelling wave resonantly tuned laser in which the modes are
presumed plane waves and the pumping is presumed uniform\cite{Haken:1975}. 
Even though these equations are isomorphic their conventional layout is
different, as shown in equations \ref{eq:lorenz1}-\ref{eq:lorenz2} and
\ref{eq:haken1}-\ref{eq:haken2}.

\begin{eqnarray}
\dot{x} & = & \sigma [y - x]                    \label{eq:lorenz1} \\
\dot{y} & = & \rho x - y - x z                                     \\
\dot{z} & = & x y - \beta z 			\label{eq:lorenz2}
\end{eqnarray}

\begin{eqnarray}
\dot{E} & = & \kappa P - \kappa E                      	\label{eq:haken1} \\
\dot{P} & = & \gamma_{\bot}[ED - P]                                       \\
\dot{D} & = & \gamma_{\|}[\lambda - 1 - D - \lambda EP]	\label{eq:haken2}
\end{eqnarray}

The variables in the Lorenz equations, namely $x$,$y$ and $z$ correspond to the
slowly varying amplitudes of the electric field $E$ and polarization $P$ and
the inversion $D$ respectively in the Lorenz-Haken equations. The parameters
are related via $\beta  = \frac{\gamma_{\|}}{\gamma_{\bot}}$, $\sigma =
\frac{\kappa}{\gamma_{\bot}}$ and $\rho = \lambda + 1$, where $\gamma_{\|}$ is
the relaxation rate of the inversion, $\gamma_{\bot}$ is the relaxation rate of
the polarization,$\kappa$ is the field relaxation rate, and $\lambda$
represents the pump power, normalized to the threshold for CW lasing.  

It was conjectured that Lorenz-like chaotic behaviour could exist in an
optically pumped far infrared ring laser. Subsequently, such chaos,
characterised by variable length trains of pulses of increasing amplitude
("spirals") was observed by Weiss and coworkers in $NH_{3}$ lasers, under
appropriate conditions\cite{Weiss:1986}. In this paper we will not discuss the
other sorts of chaos exhibited by the laser while it is under other conditions.
Detailed study has since shown that there are only a few minor differences
between the laser's Lorenz-like chaos and Lorenz Haken chaos\cite{Weiss:1995}.
One of the most obvious differences is the appearance of an anomalously large
peak nearly always observed at the start of every spiral in the
laser\cite{Weiss:1995a}. This has been attributed to deviations from the simple
model due to Doppler broadening or coherent pumping effects on the pump
transition\cite{Tang:1992}. It was also found that this effect appears if the
theory  simply accounts for the fact that the laser and the pump have
transverse structures (i.e. there exist real modes in the cavity) and are not
plane wave as the Lorenz Haken equations assume\cite{Smith:1995}. Perhaps it is
a generic effect of any small deviation from the Lorenz equations. The other
obvious difference in the behaviour of the laser is that under nearly all
circumstances the laser pulses periodically below chaos threshold and does not
make a direct transition from the steady state to chaos as the Lorenz Haken
equations predict. Although topological arguments suggest that in any
continuous transition from a fixed point to a strange attractor the attractor
should become one-dimensional, corresponding to a periodic orbit, at some
point, the transition to chaos in the Lorenz equations is a global bifurcation
with no intermediate states. Further, the model accounting for nonuniformity of
pump and electric field also shows a transition with no intermediate periodic
state. The aim of this paper is to describe the route to Lorenz like chaos in
an optically pumped far infrared ammonia laser and suggest the possible nature
of the bifurcations and hence the underlying topology of the laser attractor as
the pump power is increased from the lasing threshold to chaos.

The next section will give a short revision of the route to chaos with
increasing pump power for the Lorenz Haken equations to allow easy comparison
with the route to chaos observed in the laser, which will be described in
section \ref{route in laser}. In section \ref{detuning} the results of  section
\ref{route in laser} will be extended to include the effect of detuning in the
laser.  

\section{THE LORENZ HAKEN EQUATIONS}
\label{LH equations}

The Lorenz equations have been studied extensively over the years as a basic
illustrator of chaos and demonstration platform for various theories relating
to chaos. An extensive review of Lorenz chaos was undertaken by
Sparrow\cite{Sparrow:1982}, who subsequently showed numerically that there is
considerable complexity in the dynamics of the Lorenz equations which was not
previously  analytically explained. This included such phenomena as homoclinic
explosions which give rise to stable periodic motion, and a heteroclinic
bifurcation.  

To explain how the route to chaos in the laser differs from that of the
equations a brief revision  of the bifurcations in the equations will be given.
Below lasing threshold the origin is globally attracting. That is, all
trajectories finish at the origin if $0<\lambda+1<1$. At $\lambda>0$ a simple
bifurcation occurs which creates two stationary points at
$(\pm\sqrt{\frac{\gamma_{\bot}\lambda}{\gamma_{\|}}},
\pm\sqrt{\frac{\gamma_{\bot}\lambda}{\gamma_{\|}}},  \lambda)$ henceforth
called $C_{1}$ and $C_{2}$.  At this bifurcation the origin loses its
stability, however $C_{1}$ and $C_{2}$ are stable. The two points represent the
steady state solutions for the system in the two possible orientations that
they might find themselves, with the elctric field and polarization differing
by a $/Pi$ phase shift. At $\lambda+1=\frac{\kappa}{\gamma_{\bot}}(\frac{\kappa
+ \gamma_{\|} + 3 \gamma_{\bot}}{\kappa - \gamma_{\|} - \gamma_{\bot}})$ the
equations go through a second bifurcation in which the points $C_{1}$ and
$C_{2}$ lose their stability. The bifurcation is of the type known as a Hopf
bifurcation, which may occur in two ways. The bifurcation is "supercritical" if
each point loses its stability by expelling a stable periodic orbit.  It is
"subcritical" if they lose their stability by absorbing an unstable periodic
orbit. For values of $\lambda$ greater than the Hopf bifurcation point the
origin and both $C_{1}$ and $C_{2}$ are all unstable. There is another point on
the route to chaos which delineates an important event and this is the
occurence of a homoclinic orbit. The homoclinic orbit separates the region
between normal stable behaviour and possible  metastable chaos, associated with
an unstable periodic orbit. This can be understood by considering operation
just above lasing threshold. A trajectory started very near to the origin will
move out and, as shown in figure 1, quickly spiral into either $C_{1}$ or
$C_{2}$ depending on which the point started closer to. As $\lambda$ is
increased the trajectories around the $C_{i}$ widen until the trajectory in
both the forward and backward propagation of time moves towards the origin. 
There is therefore a homoclinic orbit associated with the origin. If $\lambda$
is increased slightly further the trajectory jumps across to spiral into the
opposite stable point. Figure 1 illustrates this point and
Sparrow\cite{Sparrow:1982} gives a more detailed explanation of how this
occurs. Increasing $\lambda$ above the homoclinic explosion, as shown by
Sparrow, the homoclinic orbit becomes an unstable periodic orbit, getting
smaller and finally getting absorbed by the stable point $C_{1}$ or $C_{2}$ at
the Hopf bifurcation.  

The implications of the existence of these orbits is best described by figure 2
where the bifurcations are placed into context with each other. When the
unstable periodic orbits decrease in size there is a possibilty for the
trajectory to oscillate between the two sides of the attractor and do so for
long periods of time. However because there are two stable attractors embedded
in an unstable attractor the trajectory must finally intersect the stable
manifolds and spiral into the stable points $C_{1}$ or $C_{2}$. So instead of
the infinitely long single trajectory as experienced in chaos there are an
infinite number of finite length trajectories. This is generally called
"preturbulence" or "metastable chaos" and refers to the fact that the dynamics
may look chaotic, but will at some finite time intersect the stable manifold
and spiral toward one of the stable points.  
 
\section{ROUTE TO CHAOS IN THE LASER}
\label{route in laser}

The experimental setup we use has been fully described by Tin Win et
al\cite{TinWin:1993}. Briefly, it is a $^{15}NH_{3}$ far infrared ring laser
which is pumped by a $^{13}CO_{2}$ laser whose pump output is attenuated by an
acousto optic modulator (AOM) of rise time 0.8 $\mu s$. To gain more insight
into the attractor properties, the laser was switched from above chaos
threshold to below chaos threshold periodically by the AOM. Each period above
chaos was about 50-100 chaotic pulses long, whereas the period below chaos
threshold was generally about 100-150 pulses in length. The time between pulses
is about 0.8 $\mu s$.  Figure 3 illustrates the time series of the Lorenz Haken
equations as a result of such a sudden decrease in $\lambda$ from just above
chaos threshold to below chaos threshold. The transition to the new state
involves two phases. Firstly there is a variable length period of chaotic
pulsations, followed by a damped oscillation. Theoretical studies of this
behaviour\cite{Yorke:1979} have shown that the length of the chaotic phase can
be characterised by an exponential distribution with a mean length dependent on
how far below chaos threshold the pump power is. A detailed experimental study
of this behaviour is being undertaken with the results to be published
elsewhere\cite{Dykstra:1997}. Figure 4 shows a typical time series for the
$^{15}NH_{3}$ far infrared laser, demonstrating the predicted behaviour. It
should be noted that the only variable which is measurable in the laser is the
intensity. This of course places some restrictions on the information which is
processable from any time series obtained. Comparison of return maps and
statistical analysis\cite{Li:1993} have shown that the chaos exhibited by the
laser, for specific parameters is very similar to Lorenz chaos. Because the
chaos is Lorenz-like it can be assumed that the dynamics above chaos threshold
must have a topologically similar structure to the Lorenz Haken equations. At
the other extreme, Tin Win et al\cite{TinWin:1993} determined the parameters
$\gamma_{\|}$, $\gamma_{\bot}$ and $\kappa$ for the $^{15}NH_{3}$ laser by
comparing the decay envelopes for transients for low pump power with the
linearised Lorenz Haken equations. These studies showed that the laser operates
in reasonably close agreement with the Lorenz Haken equations for low pump
powers.  

Figure 3 and figure 4 compare the dynamics of the experiment with what is
expected from the equations. In the parameter range where the dynamics of the
laser exhibits preturbulence its behaviour also seems to be very similar to the
Lorenz Haken equations\cite{Dykstra:1997}. In the Lorenz Haken equations, close
to but above the homoclinic explosion a trajectory started in the vicinity of
the origin should go around one of the points $C_{i}$ once and then spiral into
the other point. It is normally possible to measure only the intensity in these
experiments which obviously makes it difficult to observe a jump from one side
of the attractor to the other. One way to detect a jump is to measure the phase
of the laser radiation by heterodyne detection. We have carried out this
experiment using a second $^{15}NH_{3}$ laser (a standing wave laser operated
continuously with a few milliwatts of output power) as a local oscillator and a
Schottky barrier detector as mixer with the two beams combined at a
polyethylene beamsplitter. The output from the detector contains a homodyne
spectrum from the ring laser, extending out several megahertz, and the
heterodyne spectrum centred on the mean frequency difference between the two
lasers which was limited to about 6 $MHz$. The detector signal was digitised,
the homodyne spectrum removed numerically and quadrature field components (or
amplitude and phase) reconstructed by numerically mixing with a signal at the
mean difference frequency, in analogy with RF engineering methods. Appendix A
has a thorough explanation of the method used to extract the amplitude and
phase of the signal. This method is similar to that used by Tang et
al\cite{Tang:1992a}, but handles transients better. Examples of the results are
shown in figures 5 and 6. In both cases the laser pump power was rapidly
switched on from zero to a level giving CW output, and the transient approach
to the steady state was recorded. Figure 5 shows the intensity and phase
variation at low pump powers with an apparent slow drift in phase (a result of
a drift in local oscillator frequency relative to the ring laser) but no jump
in phase. Figure 6, however shows  the intensity and phase variation with the
pump power above the first homoclinic explosion, with a $\pi$ phase jump
clearly present between the first and second pulses.  

The only difference between the laser and the equations is the occurence of a
stable oscillation within the preturbulent regime in the laser. Instead of
finally decaying to a steady value corresponding to a fixed point, the decay is
to a stable oscillation. This oscillation is also met if the pump power is
raised slowly in a quasi-static manner, with a threshold somewhat below the
chaos threshold. The oscillation starts small at lower pump powers and grows
with increasing pump power until the laser shows no more preturbulence. When
the stable oscillation grows to be about the same size as the preturbulence,
the laser jumps into chaos and stays there. Figure 7 shows a typical graph of
percentage pump power below chaos threshold and oscillation amplitude for a
particular set of parameters.  Figure 4 shows an example of such behaviour.
When this is compared with the time series of figure 3 it is obvious that this
does not occur in the Lorenz Haken equations.  

To revise what happens as the pump power is increased;

\indent 1. \indent For pump powers below 120 $mW$ at a pressure of 38 $\mu bar$
there is no lasing

\indent 2. \indent For powers greater than this the laser makes a transient
oscillation which damps to steady state much like the Lorenz Haken equations,
moving towards one or other of the stable fixed points (Figure 5).  

\indent 3.  \indent A point started near the origin and close to Lorenz-like
chaos will exhibit a jump from one side of the attractor to the other and
subsequently collapse into steady state lasing (Figure 6).  

\indent 4. \indent For slightly higher pump powers the laser exhibits
preturbulence much like that of the Lorenz Haken equations. The differences
between them lies in the appearance of a stable oscillation in the laser which
grows in amplitude as the pump power increases (Figures 3 and 4).

So far the dynamics of the Lorenz Haken equations and the dynamics of the laser
have been described to explain their respective routes to chaos as the pump
power is increased. Both routes look quite similar except for the appearance of
the stable oscillation in the laser. A possible explanation for the occurrence
of such a stable oscillation is shown in figure 8. Starting at low pump power
the first significant event, as pump power is increased, is the appearance of
the two unstable periodic orbits around the stable points $C_{i}$ as is the
case with the Lorenz Haken equations. Increasing the pump power further the
points $C_{i}$ bifurcate in a supercritical Hopf bifurcation, each expelling a
stable periodic orbit and becoming unstable. A representative diagram showing
all the orbits for a specific pump power is shown in figure 9. Increasing the
pump power from here on results in an increasing size of the stable homoclinic
orbits. Chaos is reached when the unstable and stable homoclinic orbits
annihilate each other to leave Lorenz like chaos. The most trivial implication
of this route to chaos is that the points $C_{i}$ go unstable at lower pump
powers than they do in the Lorenz Haken equations. This also has the
implication that the range of pump powers over which preturbulence exists may
be reduced.  

\section{THE EFFECTS OF DETUNING}
\label{detuning}

During the course of the experiment it was observed that the amplitude of the
stable oscillations during preturbulent chaos is dependent upon detuning of the
laser cavity. It was found that within an extremely small range of detuning, 
preturbulence occurred with the laser relaxing down to steady state rather than
the stable oscillations. This range was far narrower than the range over which
Lorenz like chaos was observed, and so narrow that it was not possible to carry
out an exhaustive investigation. However, figure 10 shows an intensity time
series of such an event. At the moment it is not possible with our setup to
measure how far from zero detuning the laser is, no conclusions can therefore
be made as to where this region resides. It would not be surprising however to
find this region at zero detuning. This would therefore mean that there does
exist a regime where the laser chaos corresponds exactly to the Lorenz laser
case described by the Lorenz Haken equations. Such a regime has also been
observed in the $^{14}NH_{3}$ laser\cite{Weiss:1996}.  

\section{CONCLUSIONS}

It has been shown, based on measurements of the transient behaviour of the
laser, that the $^{15}NH_{3}$ far infrared laser has a route to chaos with
increasing pump power which is slightly different from that of the Lorenz Haken
equations. As with any other laser it has a minimum pump power below which no
lasing is possible. For powers greater than this the laser makes a transient
oscillation which damps to steady state much like the Lorenz Haken equations.
Using heterodyne techniques to see the field there was found to be a pump power
above which the laser exhibits a jump from one side of the attractor to the
other where the field subsequently collapsed to steady state lasing. At
slightly higher pump powers the laser was found to exhibit preturbulence
leading to a stable oscillation which grows in amplitude with increasing pump
power. Finally, above a certain power Lorenz like chaos is observed.  

The route to chaos, as observed in our laser, has been conjectured to involve
an extra stable periodic orbit which annihilates with the unstable periodic
orbit to leave Lorenz like chaos.  

It was also shown that for an extremely small range of detunings the laser will
show preturbulence without the occurence of the stable oscillations. These
phenomena will provide an excellent test for improved theoretical models of the
laser.  
 
\section{Appendix A}

In the heterodyne measurement of the chaotic laser, the signal from the chaotic
laser and the signal from the local oscillator are coupled together by a beam
splitter. The mixed signals are then detected by a Schottky barrier diode. The
electric field of the chaotic laser and the local oscillator are:
\begin{eqnarray}
E_{1}(t) & = & A_{1}(t)\sin[\omega_{1} t+\phi(t)] \\
E_{2}(t) & = & A_{2}\sin(\omega_{2} t)
\end{eqnarray}
where  $A_{1}$, $\omega_{1}$ and $\phi$ are the amplitude, the ``optical"
frequency and the phase of the electrical field of the ring laser respectively.
$A_{2}$ and  $\omega_{2}$ are the amplitude and the ``optical" frequency of the
local oscillator respectively. $A_{1}$ and $\phi$ are time dependent. $A_{2}$,
$\omega_{1}$ and $\omega_{2}$ are constant.

The response of the Schottky diode is~\cite{barber:1969}
\begin{equation}
I(t) \propto  e^{\alpha V}=e^{\beta ( E_{1}+E_{2})}
=1+\beta( E_{1}+E_{2})+\frac{1}{2}\beta^{2}( E_{1}+E_{2}) ^{2}+\ldots
\end{equation}

where $\alpha$ and $\beta$ are constants.

Because the rest of the electronics  responds only to frequencies much lower
than the ``optical" frequencies $\omega_{1}$ and $\omega_{2}$, and a bias
voltage is applied to the Schottky diode to make it favor the square term,  in
the output from the Schottky detector only the square term is important. So for
$\Delta t \gg 1/\omega_{1},1/\omega_{2}$ ($\Delta t$ is the response time of
the Schottky diode) the output from the Schottky diode is
\begin{eqnarray}
<I>_{\Delta t} & \propto & < (E_{1}+E_{2})^{2}> \nonumber \\
 & = & <A_{1}^{2}\sin^{2}(\omega_{1} t + \phi)+A_{2}^{2}\sin^{2}(\omega_{2} t)
+2A_{1}A_{2}\sin(\omega_{1} t +\phi)\sin(\omega_{2} t)> \nonumber \\
 & \propto   &\underbrace{A_{1}^{2}+A_{2}^{2}}_{homodyne\: part}+
\underbrace{2A_{1}A_{2}\cos[(\omega_{1}-\omega_{2})t+\phi]}_{heterodyne\: part}
\end{eqnarray}

The output from the Schottky diode is a mixture of the desired heterodyne
signal carried by the beat between the chaotic laser and the local oscillator
with an unwanted homodyne component. The time variation of $A_{1}$ and $\phi$
lead to a finite width for both the homodyne and heterodyne spectra in the
frequency domain. It was ensured that the spectra did not overlap so that the
homodyne and heterodyne signals could be separated by using simple filters. If
they overlapped a more complicated detection setup, for instance using two
Schottky diodes to detect the homodyne signal and the mixed signal
respectively, would be necessary.  

In the heterodyne signal, $A_{1}$ and $\phi$ are carried by the beat between
the chaotic laser and the local oscillator. The carrier signal must be removed
from the heterodyne signal in order to reconstruct the chaotic laser field. A
Fast Fourier Transformation (FFT) was applied to the experimental record to
obtain the signal spectrum and determine the cutoff frequencies for a low and
high frequency filter. These filters were subsequently used on the original
experimental record to separate the homodyne and heterodyne signals. The
homodyne signal was constructed directly from the low pass filtered portion of
the signal. The heterodyne time series was separately mixed with a sinusoidal
and a cosinusoidal signal of the beat frequency and low pass filtered in order
to reconstruct the two orthogonal components of the heterodyne time series and
hence the field components and phase.

Tests on synthetic data show this method is reliable for in analyzing transient
signals, as it is able to reconstruct the original field amplitude and phase
perfectly. By contrast, the signal processing method used by Tang et
al\cite{Tang:1992a}, which is based on maniopulation of Fourier transforms, is
able to reconstruct periodic data reasonably well, but it introduces errors in
the amplitude and the phase when the signals are transient.

\pagebreak

\section*{Figure Captions}

\vspace*{0.125in} \noindent Figure 1: Trajectories of the Lorenz Haken laser equations at pump powers, $\lambda$ equal to 1.0 and 2.0 below the first homoclinic explosion, $\approx$3.0351841193 for the homoclinic orbit, and 4.0 above the homoclinic explosion. Parameter values of $\gamma_{\|} = 1/8$ , $\gamma_{\bot} = 1/2$ and $\kappa = 1$ are appropriate for the $^{15}NH_{3}$ laser and give a chaos threshold of 13. The circles indicate the stable points towards which the trajectories will gravitate.     

\vspace*{0.125in} \noindent Figure 2: The bifurcation diagram for the Lorenz Haken laser equations. The $x=y$ axis symbolises the symmetry along that plane of the equations.  


\vspace*{0.125in} \noindent Figure 3: Intensity time trace of preturbulence in the Lorenz Haken laser equations at $\gamma_{\|} = 1/8$ , $\gamma_{\bot} = 1/2$ and $\kappa = 1$ resulting from the fast switch down from $\lambda = 14$ above the chaos threshold to $\lambda = 10$ below the chaos threshold.   

\vspace*{0.125in} \noindent Figure 4: Typical intensity time trace of preturbulence in the $^{15}NH_{3}$ laser as the laser is switched from slightly above the 5.85 $W$ chaos threshold to a factor of 0.1 below chaos threshold.     

\vspace*{0.125in} \noindent Figure 5: Phase and intensity evolution of the $^{15}NH_{3}$ laser. The dashed line shows the decaying pulses of the intensity and the solid line the phase of the light. The phase shows a slow drift as the pulses decay indicating the absence of a jump from one side of the attractor ot the other. The pump power was a factor of 0.7 below chaos threshold.    
  
\vspace*{0.125in} \noindent Figure 6: Phase and intensity evolution of the $^{15}NH_{3}$ laser at switch on. The dashed line shows the decaying pulses of the intensity and the solid line the phase of the light. The phase shows a distinctive jump of $\pi$ after the first pulse to indicate a jump from one side of the attractor to the other. The pump power was a factor of 0.5 below chaos threshold.
  
\vspace*{0.125in} \noindent Figure 7: Amplitude of the intensity oscillations observed in the $^{15}NH_{3}$ laser as a function of pump power.    
  
\vspace*{0.125in} \noindent Figure 8: The bifurcation diagram of the $^{15}NH_{3}$ laser. The difference between this diagram and that for the Lorenz Haken equations shown in figure 2 is the presence of a supercritical Hopf bifurcation within, and which finally annihilates, the already existing homoclinic orbit.   
 
\vspace*{0.125in} \noindent Figure 9: Representation of the stable and unstable orbits relative to their respective fixed points. The format of the figure is as for figure 2.  

\vspace*{0.125in} \noindent Figure 10: Preturbulence in the the $^{15}NH_{3}$ laser for the detuning range where the preturbulence settles down to a steady state. The laser is switched from slightly above the 5.85 $W$ chaos threshold to a factor of 0.05 below chaos threshold. 

  
\pagebreak
\psfig{file=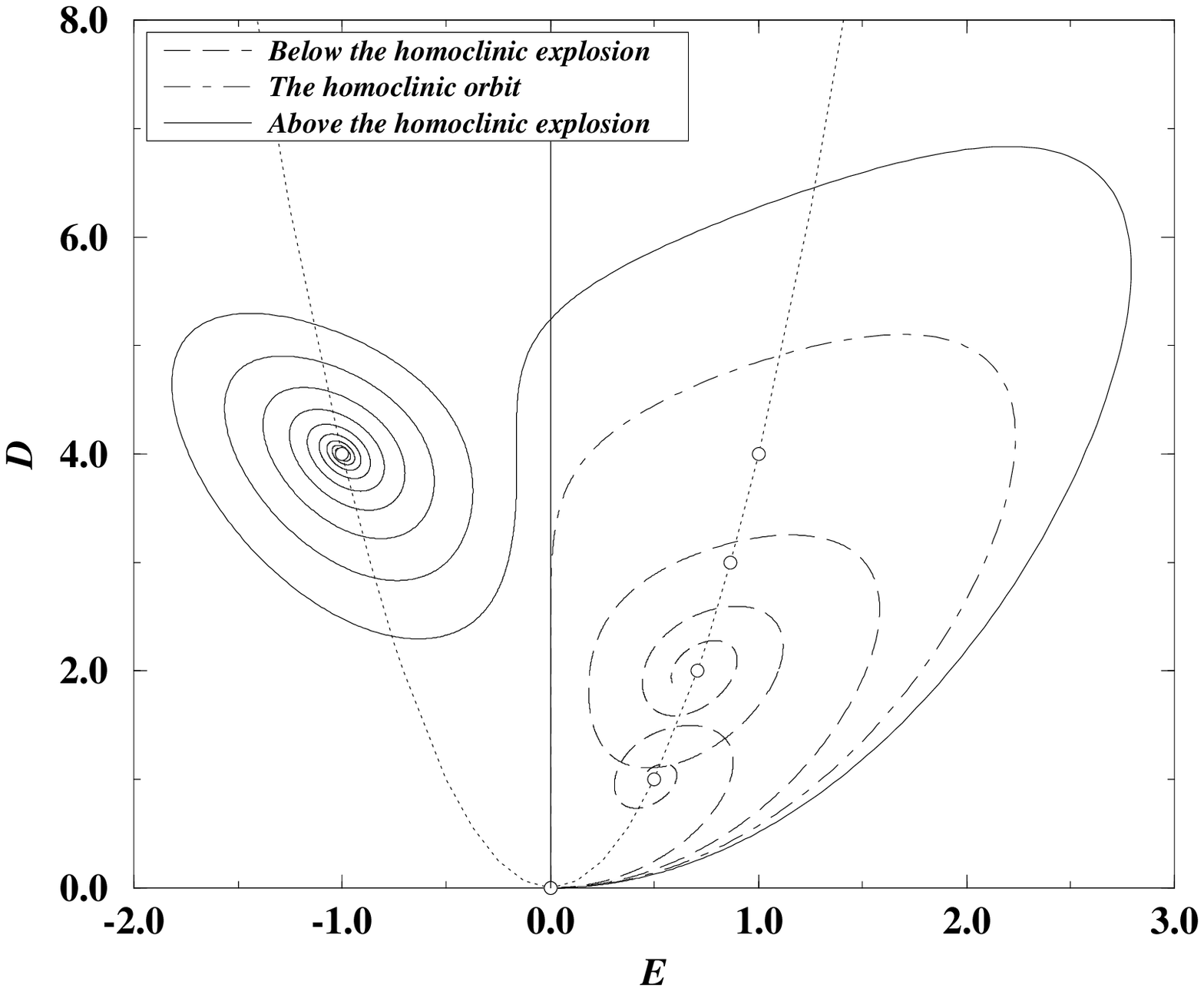,width=6in}
\pagebreak
\psfig{file=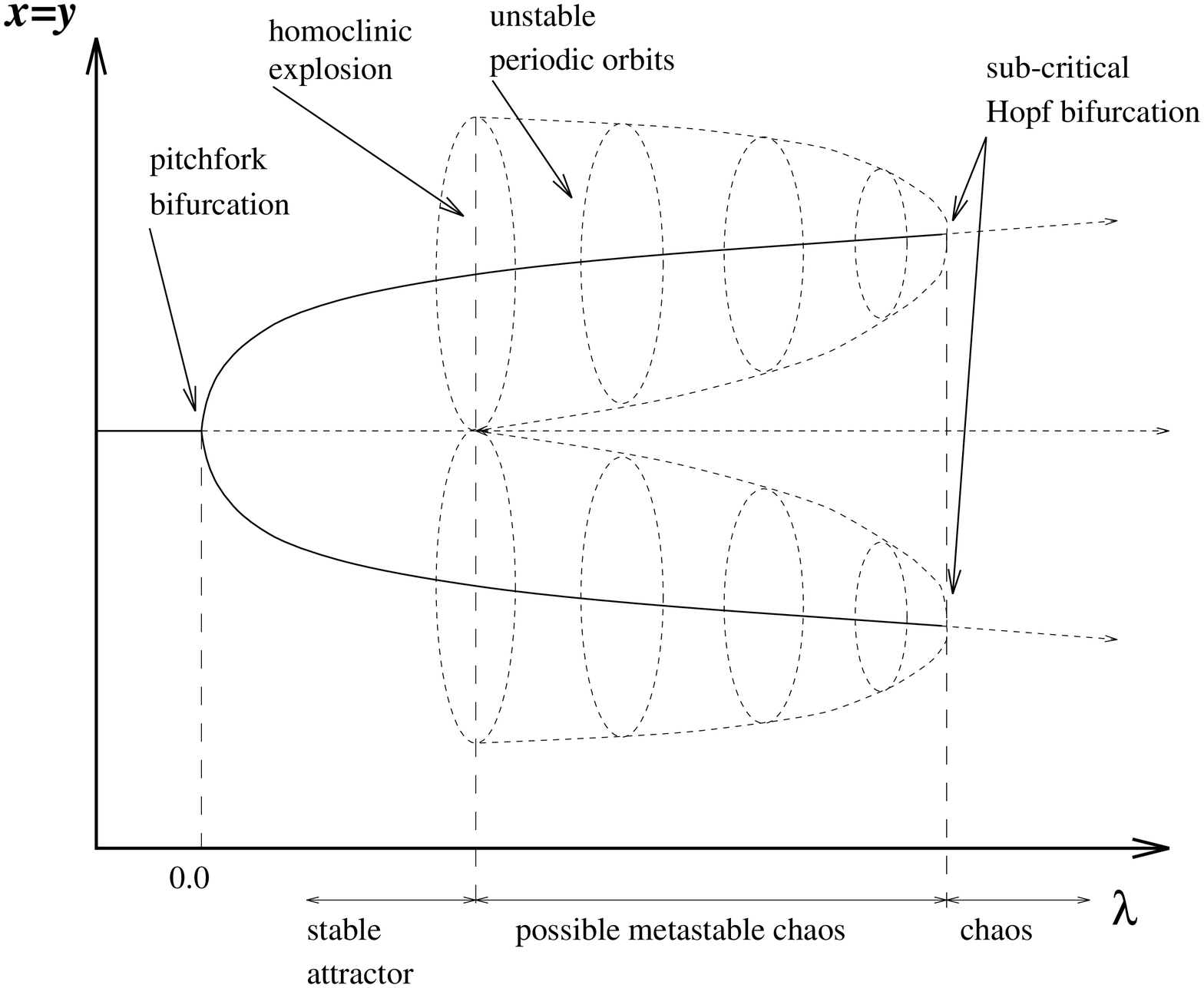,width=6in}
\pagebreak
\psfig{file=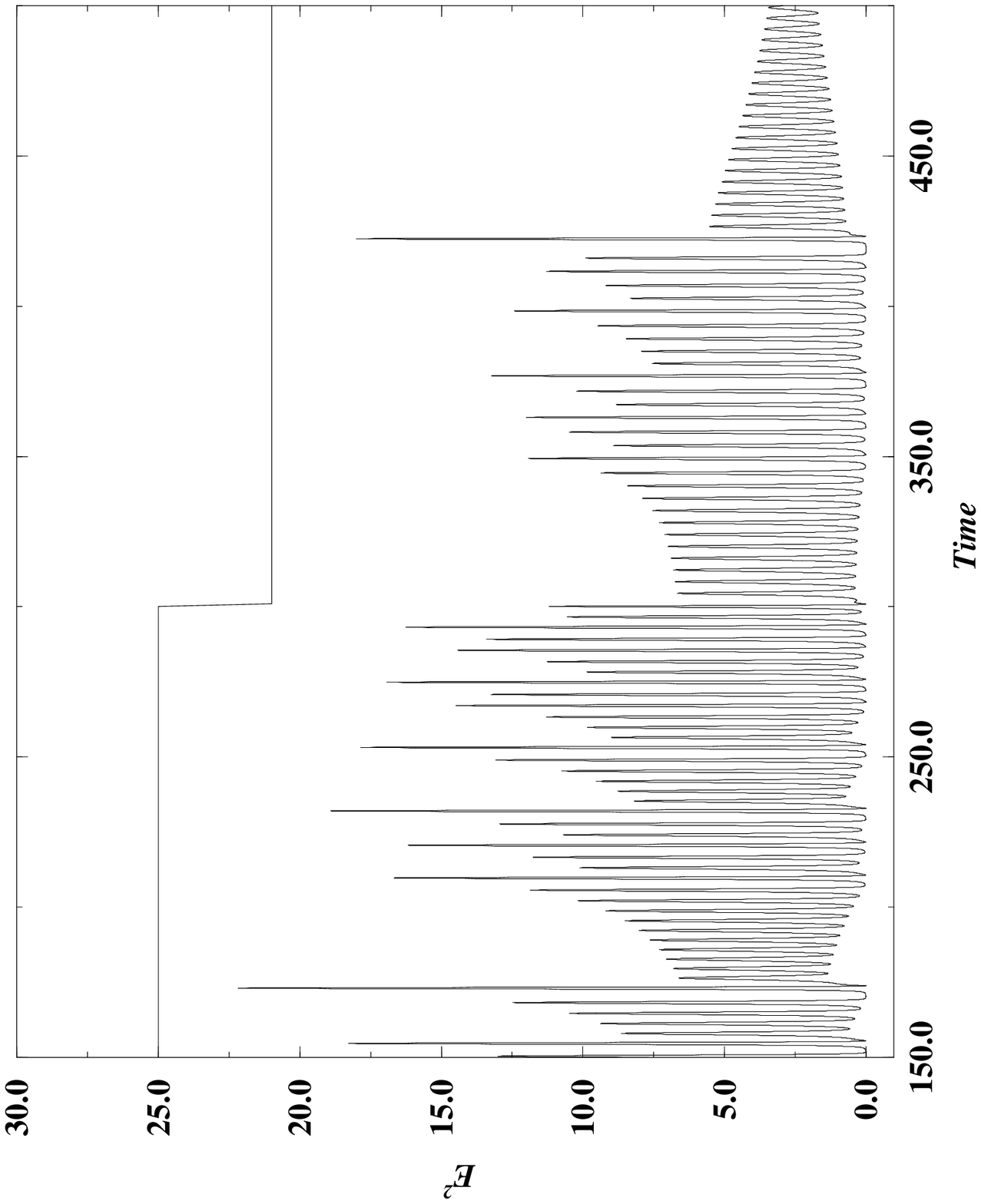,width=6in,height=8.5in}
\pagebreak
\psfig{file=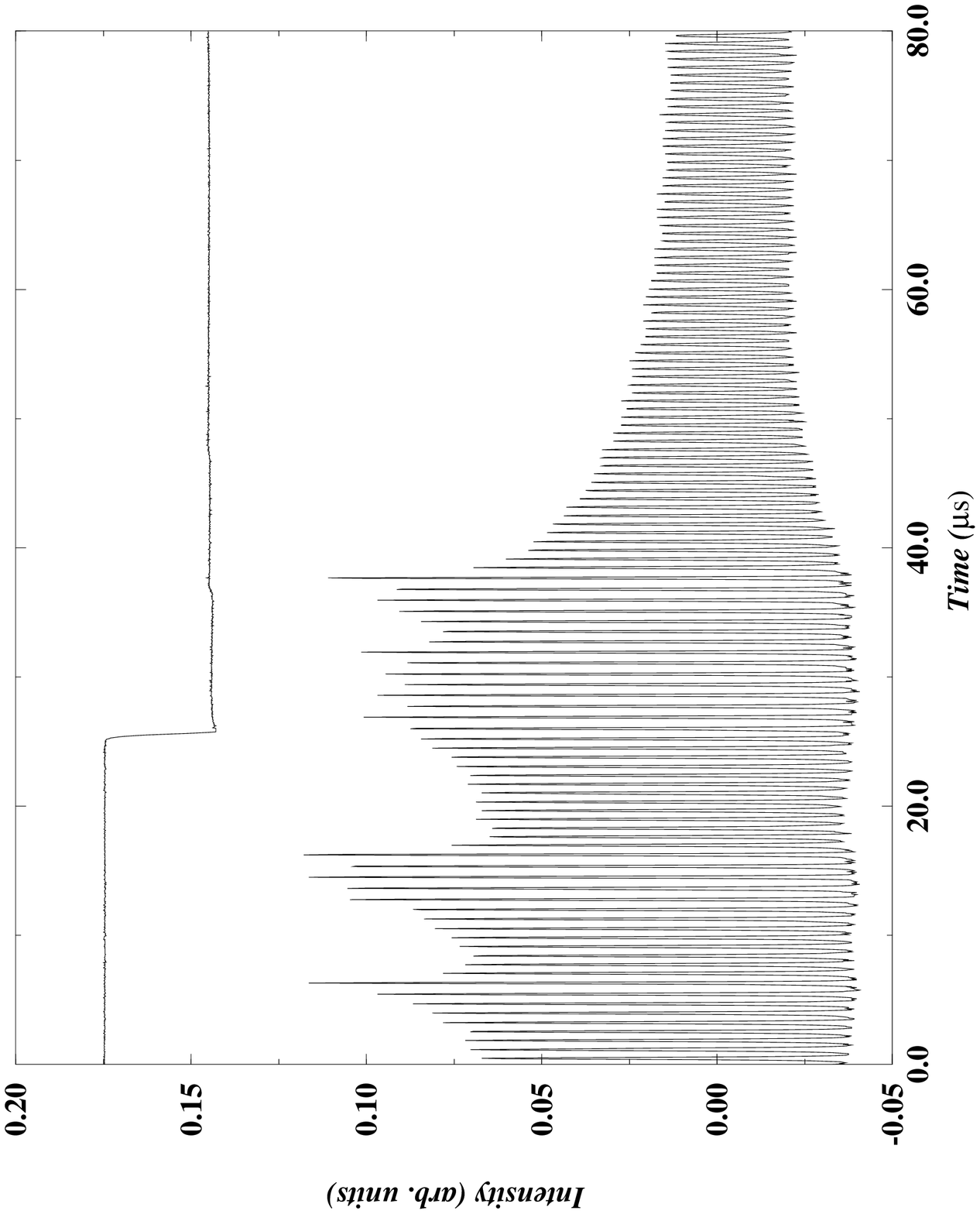,width=6in,height=8.5in}
\pagebreak
\psfig{file=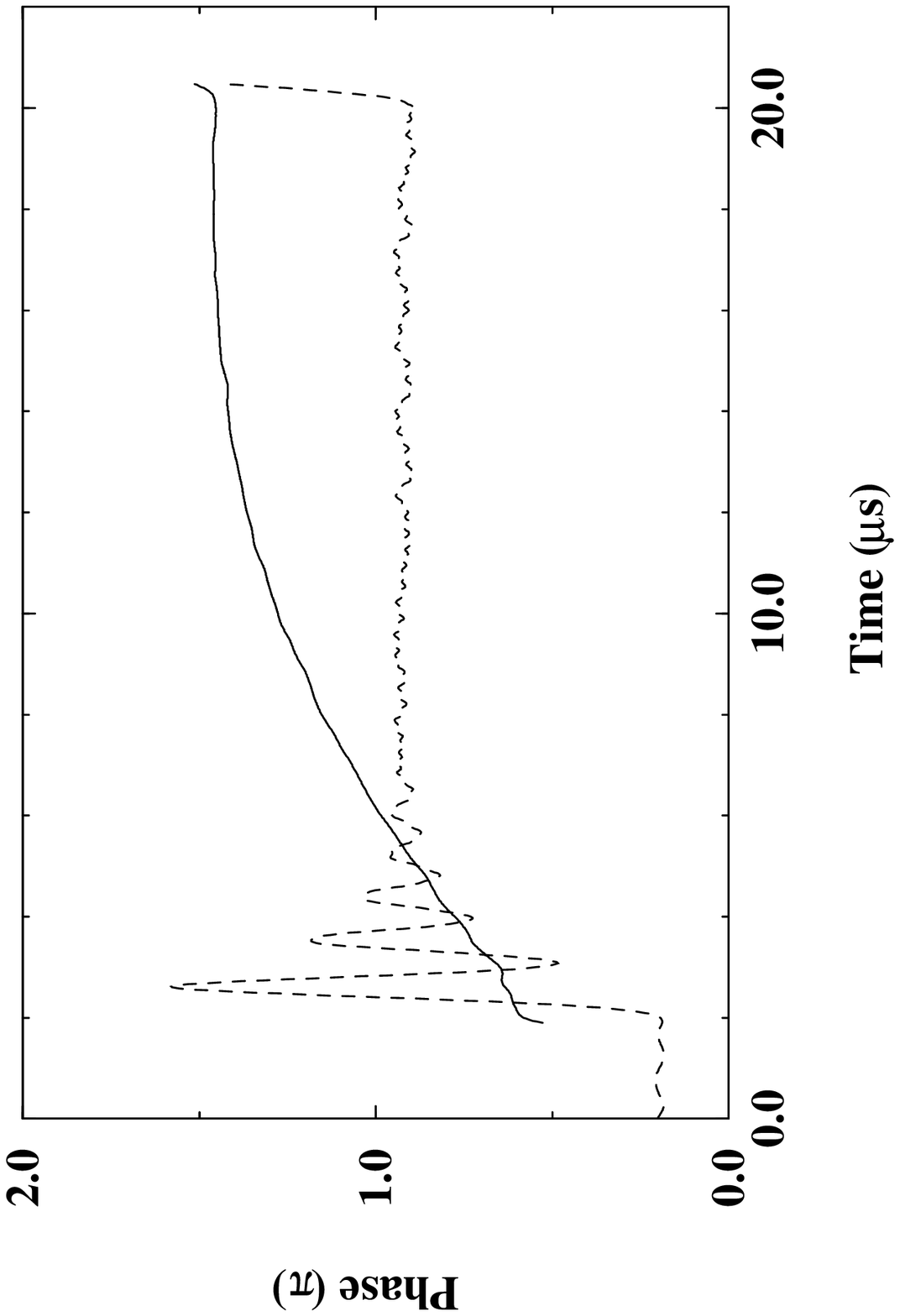,width=6in}
\pagebreak
\psfig{file=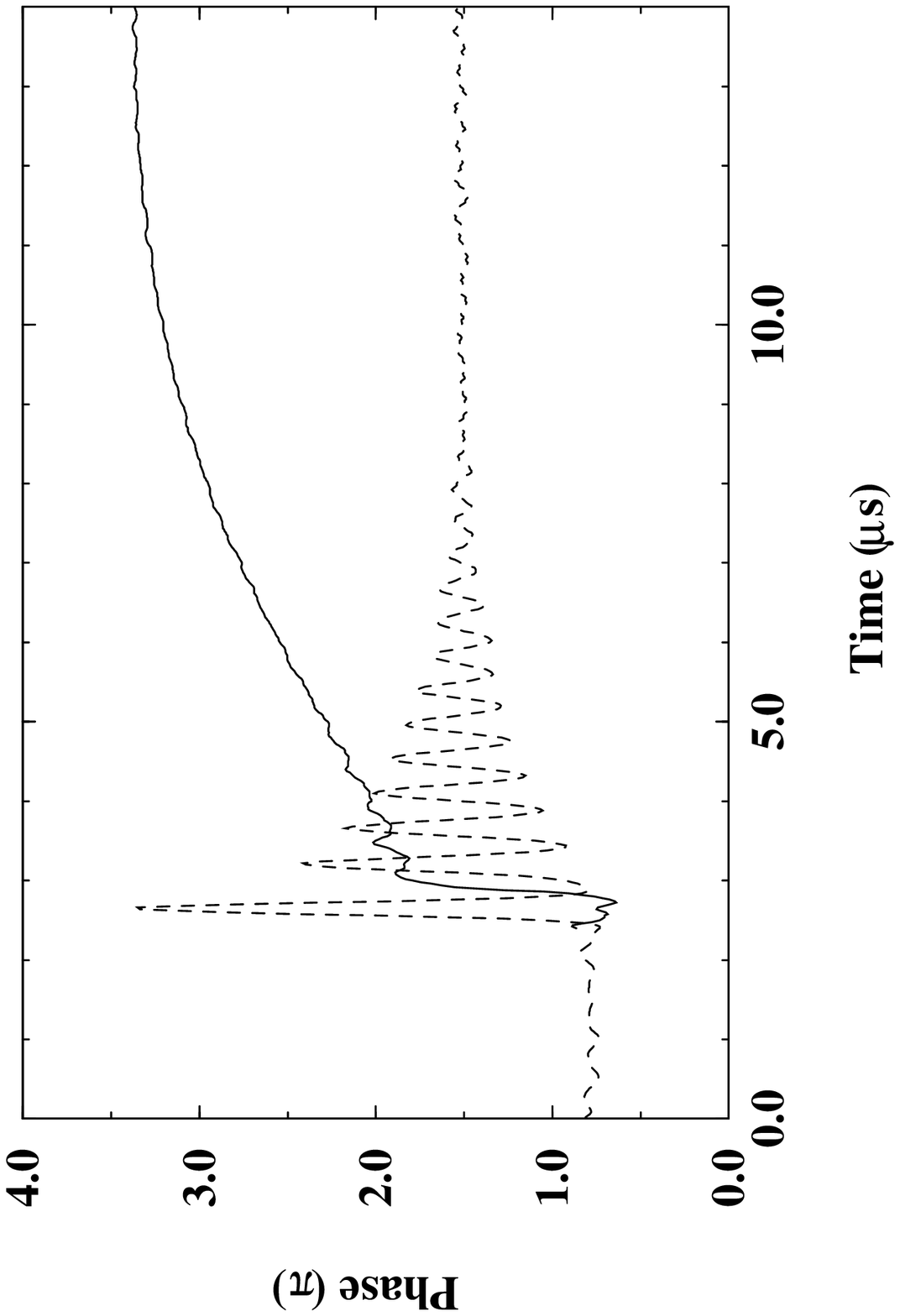,width=6in}
\pagebreak
\psfig{file=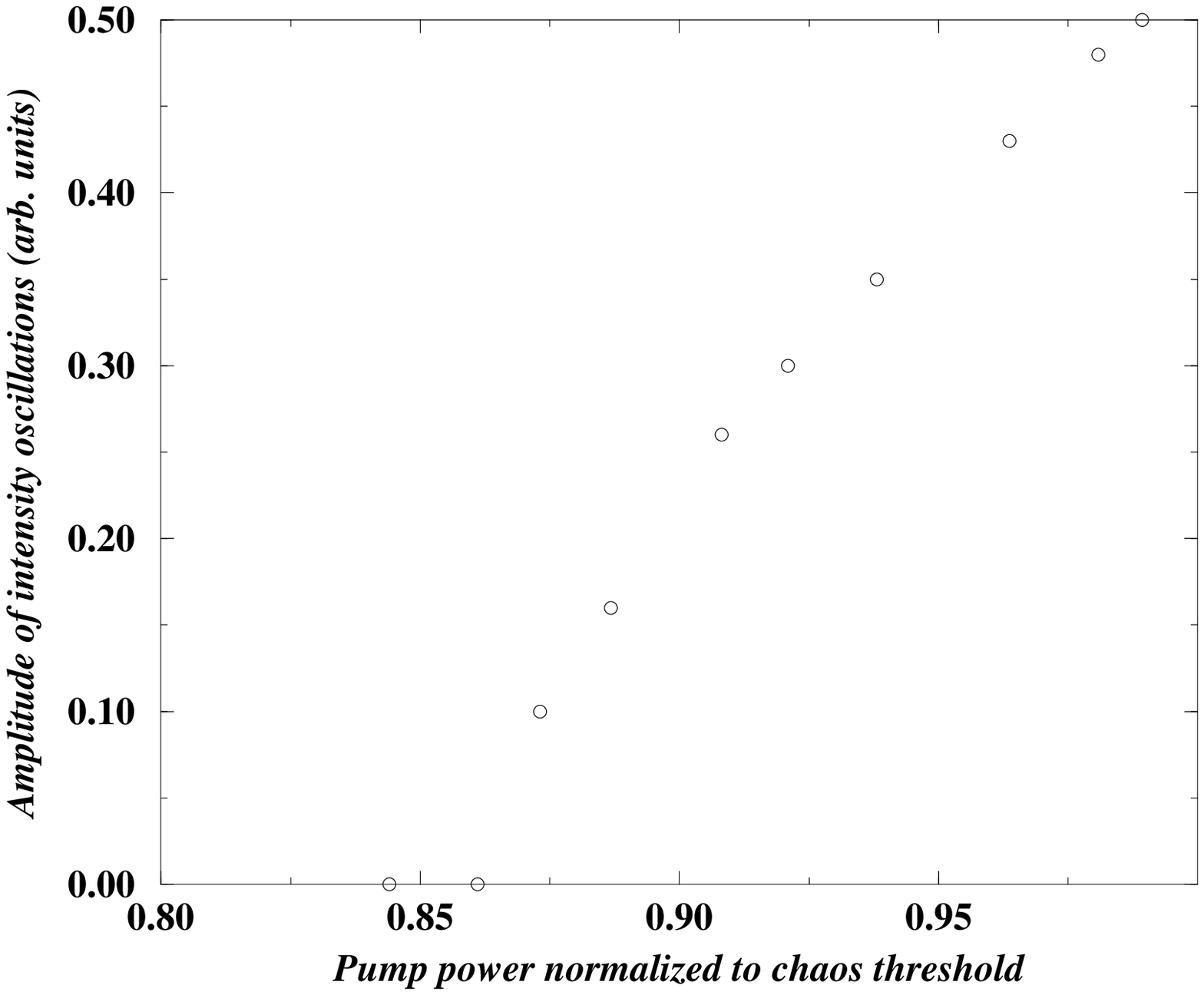,width=6in}
\pagebreak
\psfig{file=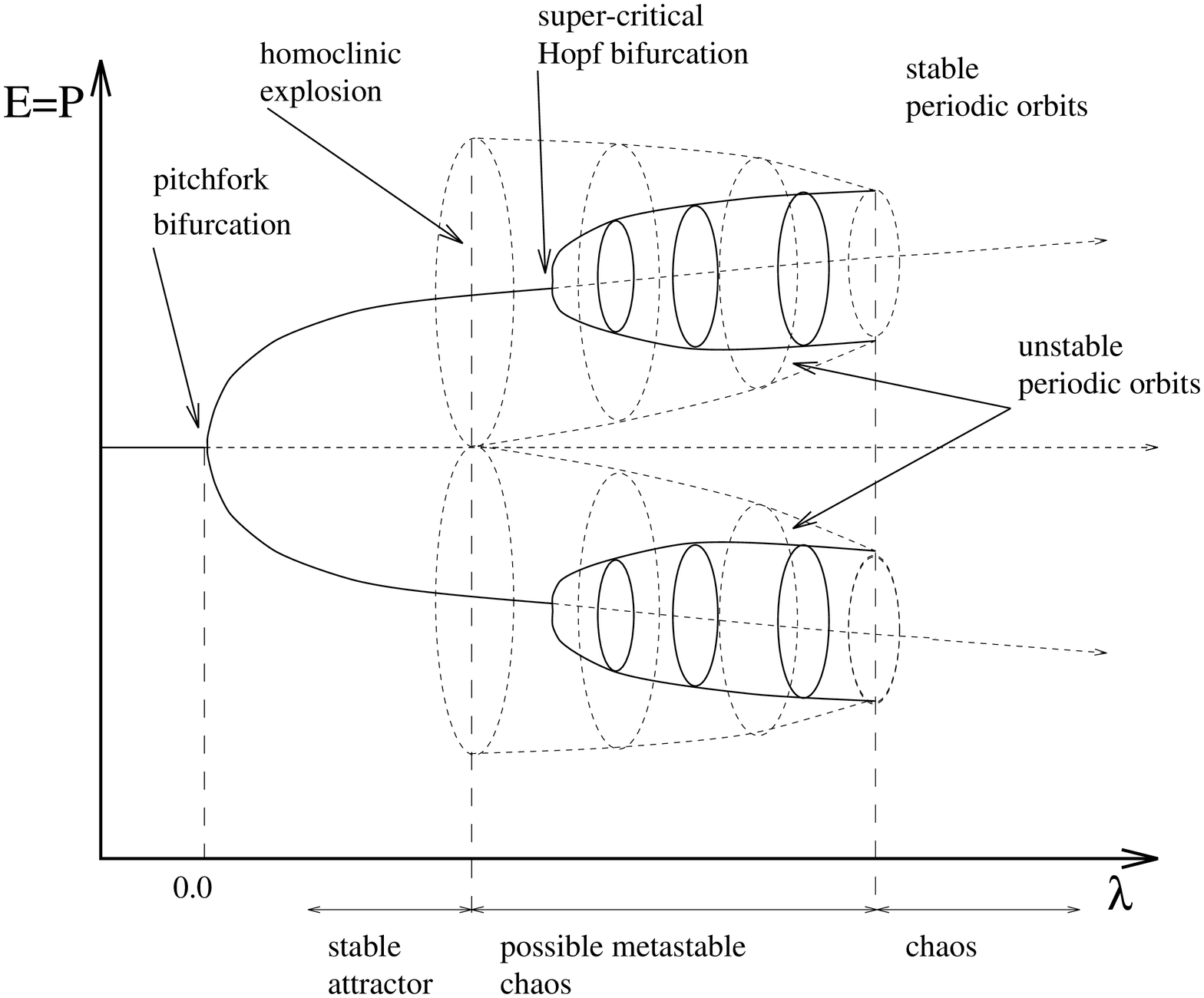,width=6in}
\pagebreak
\psfig{file=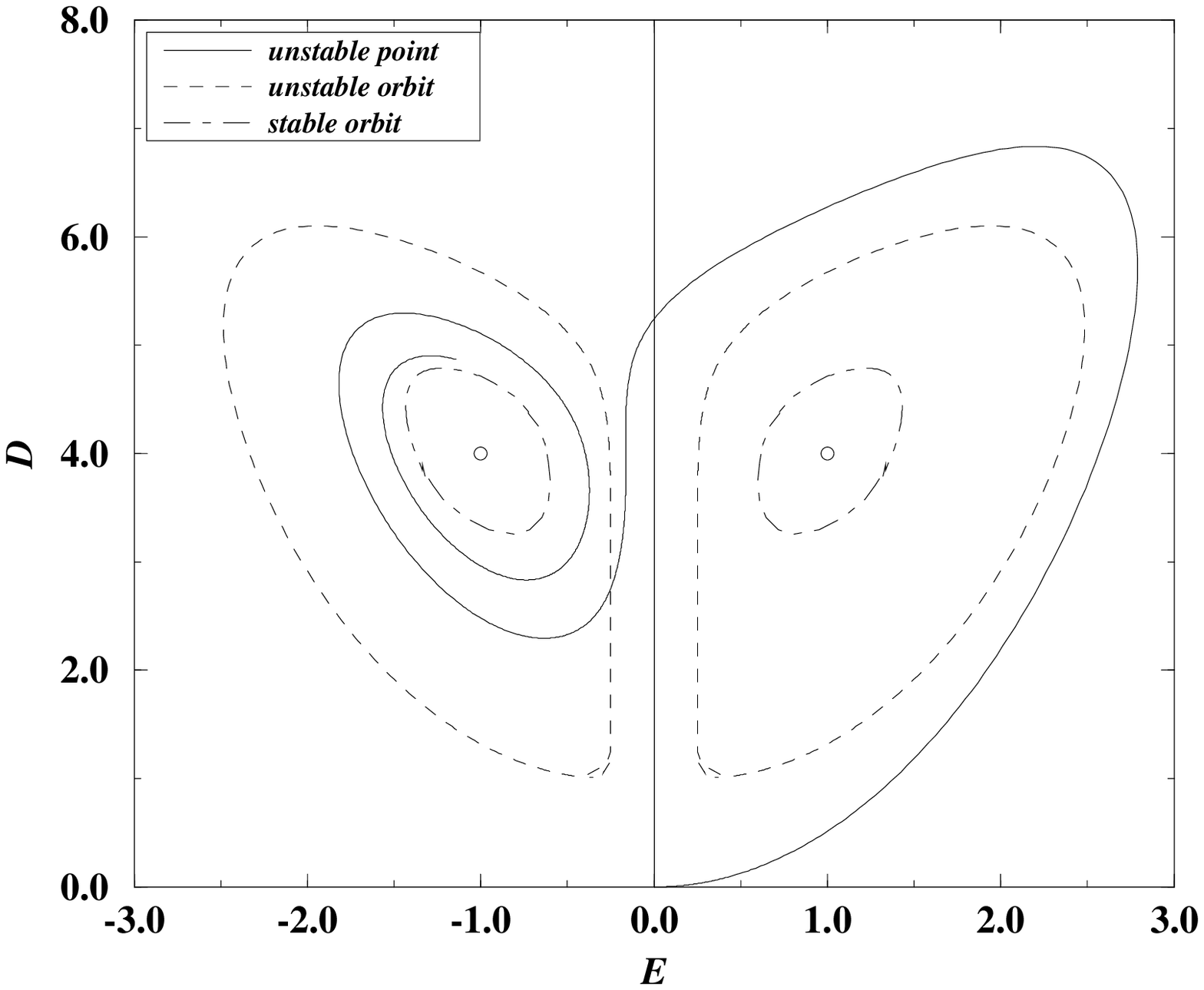,width=6in}
\pagebreak
\psfig{file=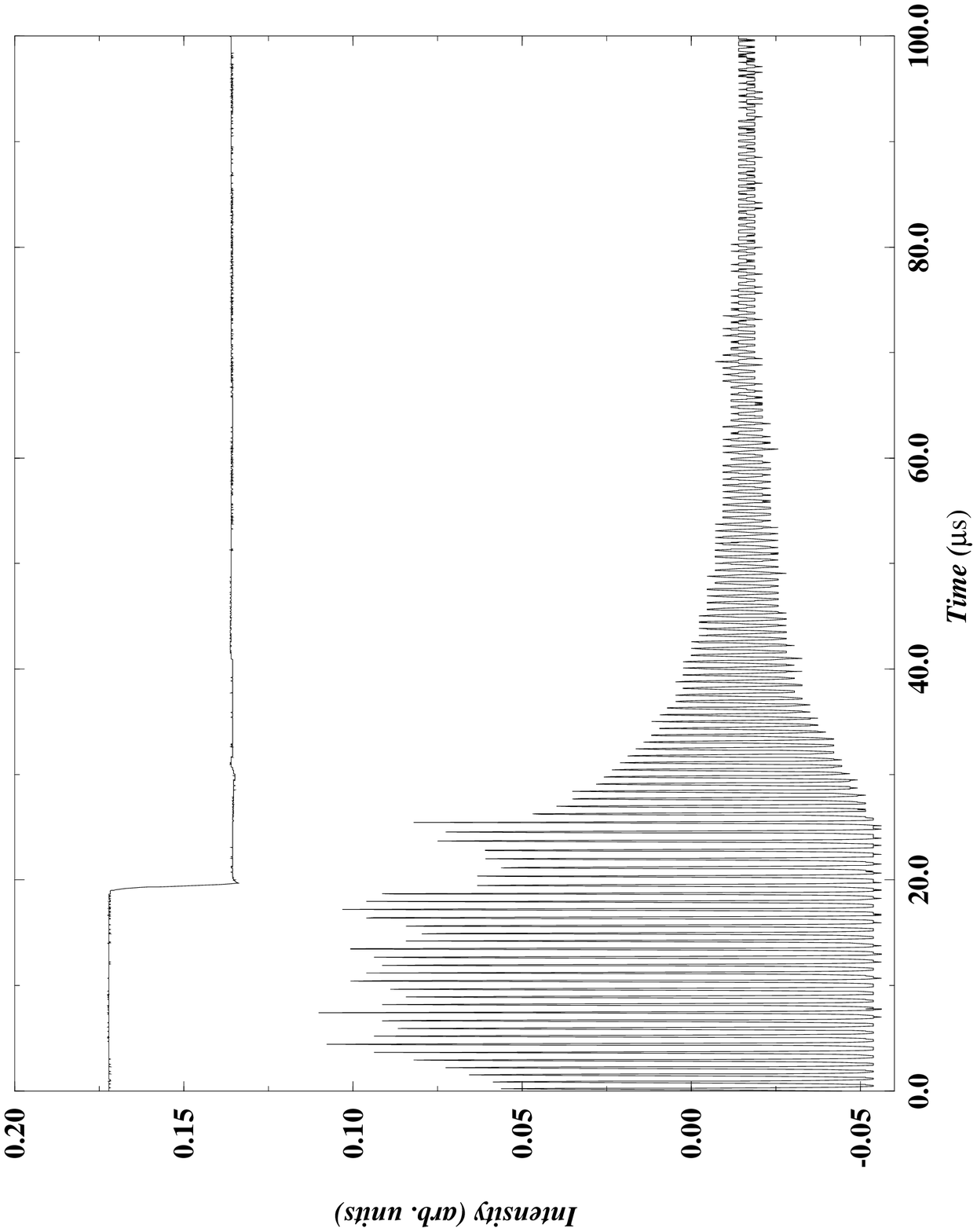,width=6in,height=8.5in}


\begin{references}
\bibitem[1]{Lorenz:1963}E.N. Lorenz, J. Atmos. Sci. {\bf 20}, 130 (1963).
\bibitem[2]{Haken:1975}H. Haken, \pl {\bf 53A}, 77 (1975).
\bibitem[3]{Weiss:1986}C.O. Weiss and J. Brock, \prl {\bf 57}, 2804 (1986).
\bibitem[4]{Weiss:1995}C.O. Weiss, R. Vilaseca, N.B. Abraham, R. Corbal\'{a}n,
E. Rold\'{a}n, G.J. de Valc\'{a}rcel, J. Pujol, U. H\"{u}bner and D.Y. Tang,
\ap B {\bf 61}, 223 (1995).
\bibitem[5]{Weiss:1995a}C.O. Weiss, U. H\"{u}bner, N.B. Abraham and D.Y. Tang,
Infrared Phys. Technol. {\bf 36}, 489 (1995).
\bibitem[6]{Tang:1992}D.Y. Tang and C.O. Weiss, \ap B {\bf 54}, 2548 (1992) 

\bibitem[7]{Smith:1995}C.P. Smith and R. Dykstra, \oc {\bf 117} 107 (1995), and C.P. Smith and R. Dykstra, \oc accepted (1996).

\bibitem[8]{Sparrow:1982}C. Sparrow, {\em The Lorenz Equations: Bifurcations, Chaos, and Strange Attractors} (Springer-Verlag, New York, 1982).

\bibitem[9]{TinWin:1993}Tin Win, M.Y. Li, J.T. Malos and N.R. Heckenberg, \oc {\bf 103}, 479 (1993).

\bibitem[10]{Yorke:1979}J.A. Yorke and E.D. Yorke, J. Stat. Phys. {\bf 21}, 263 (1979).

\bibitem[11]{Dykstra:1997}R. Dykstra, J.T. Malos and N.R. Heckenberg, \pra {\bf 56} (1997).

\bibitem[12]{Li:1993}M.Y. Li and N.R. Heckenberg, \oc {\bf 108} 104 (1993).
\bibitem[13]{Li:1995}M.Y. Li, {\em Chaos in Single and Multimode Lasers}, (PhD
Thesis, University of Queensland, 1995).
\bibitem[14]{Tang:1992a}D.Y. Tang, M.Y. Li  and C.O. Weiss, \pra {\bf 44}, 676
(1992).
\bibitem[15]{Weiss:1996}C.O. Weiss, private communication
\bibitem[16]{barber:1969}M.R. Barber {\em Microwave Semiconductor Devices and
their Circuit Applications} (McGraw-Hill, New York, 1969).

\end{references}
\end{document}